\documentclass[iop,apjl]{emulateapj}

\usepackage{natbib}
\slugcomment{To appear in The Astrophysical Journal}

\shorttitle{LITHIUM ABUNDANCES IN NEARBY FGK DWARF AND SUBGIANT STARS}
\shortauthors{RAM\'IREZ ET AL.}

\newcommand{\ali}{A_\mathrm{Li}}
\newcommand{\feh}{\mathrm{[Fe/H]}}
\newcommand{\teff}{T_\mathrm{eff}}
\newcommand{\logg}{\log g}

\newcommand{\fei}{Fe\,\textsc{i}}
\newcommand{\feii}{Fe\,\textsc{ii}}

\begin{document}

\title{LITHIUM ABUNDANCES IN NEARBY FGK DWARF AND SUBGIANT STARS: \\ INTERNAL DESTRUCTION, GALACTIC CHEMICAL EVOLUTION, AND EXOPLANETS}

\author{I.\,Ram\'irez\altaffilmark{1},
        J.\,R.\,Fish\altaffilmark{2,3},
        D.\,L.\,Lambert\altaffilmark{1}, and
        C.\,Allende Prieto\altaffilmark{4,5}
	}
\altaffiltext{1}{McDonald Observatory and Department of Astronomy,
                     University of Texas at Austin, 1 University Station, C1400
                     Austin, Texas 78712-0259, USA}
\altaffiltext{2}{The Observatories of the Carnegie Institution for Science,
                     813 Santa Barbara Street, Pasadena, California 91101, USA}
\altaffiltext{3}{Harvey Mudd College,
                     301 Platt Boulevard, Claremont, California 91711, USA}
\altaffiltext{4}{Instituto de Astrof\'isica de Canarias,
                     38205, La Laguna, Tenerife, Spain}
\altaffiltext{5}{Departamento de Astrof\'isica, Universidad de La Laguna,
                     38206 La Laguna, Tenerife, Spain}

\begin{abstract}
We derive atmospheric parameters and lithium abundances for 671 stars and include our measurements in a literature compilation of 1381 dwarf and subgiant stars. First, a ``lithium desert" in the effective temperature ($\teff$) versus lithium abundance ($\ali$) plane is observed such that no stars with $\teff\simeq6075$\,K and $\ali\simeq1.8$ are found. We speculate that most of the stars on the low $\ali$ side of the desert have experienced a short-lived period of severe surface lithium destruction as main-sequence or subgiant stars. Next, we search for differences in the lithium content of thin-disk and thick-disk stars, but we find that internal processes have erased from the stellar photospheres their possibly different histories of lithium enrichment. Nevertheless, we note that the maximum lithium abundance of thick-disk stars is nearly constant from $\feh=-1.0$ to $-0.1$, at a value that is similar to that measured in very metal-poor halo stars ($\ali\simeq2.2$). Finally, differences in the lithium abundance distribution of known planet-host stars relative to otherwise ordinary stars appear when restricting the samples to narrow ranges of $\teff$ or mass, but they are fully explained by age and metallicity biases. We confirm the lack of a connection between low lithium abundance and planets. However, we find that no low $\ali$ planet-hosts are found in the desert $\teff$ window. Provided that subtle sample biases are not responsible for this observation, this suggests that the presence of gas giant planets inhibit the mechanism responsible for the lithium desert.
\end{abstract}


\section{INTRODUCTION}

The current lithium abundance in the solar photosphere is over two orders of magnitude lower than the meteoritic value \cite[e.g.,][]{asplund09:review}, implying significant lithium depletion since the solar system formed. For many decades, this observational fact has posed a severe problem for solar interior models. The temperatures needed for astration of lithium are found just below the base of the present-day solar convective zone. Therefore, according to standard models, there should be no surface lithium depletion during the main-sequence life of stars like the Sun \cite[e.g.,][]{dantona94}. Yet, in the solar neighborhood, lithium abundances have been found to vary by more than two orders of magnitude among stars with similar atmospheric parameters, in particular near the solar values \cite[e.g.,][]{lambert04}. Several mechanisms have been proposed to explain the observed lithium abundances in solar-type stars, including rotational mixing \cite[e.g.,][]{pinsonneault10}, gravity waves \cite[e.g.,][]{charbonnel05}, and atomic diffusion \cite[e.g.,][]{michaud86}, but a fully consistent picture remains elusive.

Another important problem for models of lithium depletion is the observation of a lithium chasm in the effective temperature ($\teff$) range between about 6500 and 6850\,K in stars in young open clusters, particularly the Hyades \cite[e.g.,][]{boesgaard86,soderblom93}. Within this narrow temperature range, whose center corresponds to about $1.4\,M_\odot$ in the case of the Hyades, stars have significantly lower lithium abundances than those immediately outside of that window. That this so-called ``lithium dip'' is developed in main-sequence rather than pre-main sequence stars is demonstrated by the fact that it is not seen among stars of the youngest open clusters \citep{boesgaard88,balachandran11}. The lithium dip is, as expected, seen in field main-sequence and subgiant stars \citep{balachandran90,lambert04} where the dip moves to lower mass with lower metallicity \citep{balachandran95}. Attempts have been made also to explain the lithium dip in young open clusters \cite[e.g.,][]{talon98,mendes99}, but the overall picture is still far from being understood completely.

The complex behavior of lithium abundances observed in FGK stars and the poorly constrained mechanisms of lithium depletion inside the stars make it difficult to use stellar lithium abundances to constrain Galactic chemical evolution models \cite[e.g.,][]{romano99}. Sites of lithium production have been identified, but their relative importance for the lithium enrichment of the interstellar medium is still somewhat debatable \cite[e.g.,][]{travaglio01}. To the best of our knowledge, there has not been a study of lithium abundance evolution in different local disk stellar populations.

Studies of samples of stars with lithium abundances homogeneously derived (within each study) have been published in the past few years. Analysis of these samples has allowed a better understanding of lithium depletion, but an overall picture still needs to be drawn from the wealth of available data.

Recently, \cite{ramirez07} have extended their high resolution spectroscopic study of nearby stars from about 500 to 800 objects (Ram\'irez, Allende Prieto, \& Lambert, in preparation, hereafter RAL). Their data are being used to study oxygen abundances in local stellar populations, but their spectra cover the 6708\,\AA\ region, which contains the lithium doublet that is commonly used to derive lithium abundances. In this paper, we use RAL's dataset as a starting point to construct the largest available catalog of lithium abundances in nearby FGK dwarf and subgiant stars. The catalog is constructed using also data from the literature, but RAL's stellar parameters and the lithium abundances derived here are used to normalize all data sets to a common scale.

Given the nature of the RAL work, our sample is suited to study systematic differences in the chemical evolution of solar neighborhood stellar populations. At least two distinct populations of stars, with different spatial distributions and kinematical properties, co-exist in the Galactic disk, namely the thin disk and the thick disk \cite[e.g.,][]{gilmore83,soubiran93}. Moreover, it is known that these two populations differ in chemical composition, particularly in the $\alpha$-elements and oxygen \cite[e.g.,][]{bensby05,reddy06,ramirez07}. Models of Galactic formation and evolution must be able to explain these chemical and kinematical differences. As mentioned before, we are not aware of any prior analysis of differences in lithium abundances between the thin and thick disks, which can be investigated by studying stars that have experienced the least amount of lithium depletion. RAL's data set includes also spectra of a number of old halo stars, which we can use to shed new light on the problem of the primordial, or cosmological, lithium abundance.

Some authors \cite[e.g.,][]{gonzalez08,israelian09} claim to have revealed a connection between stellar lithium abundance and the presence of exoplanets in Sun-like stars, although others have not found evidence for this correlation \cite[e.g.,][]{ryan00,baumann10}. The mechanisms proposed to explain how the presence of planets could affect lithium depletion are planetary migration, which affects the evolution of the angular momentum of the star, as suggested by \cite{castro08}, and interactions between the protoplanetary disk and the star, which determine the degree of differential rotation between the radiative core and the convective envelope, therefore having an important impact on rotational mixing \cite[e.g.,][]{bouvier08}. In both cases, enhanced lithium depletion is expected for stars hosting planets. Lithium abundances in stellar photospheres could, therefore, be of interest also for exoplanet research. However, sample biases must be clearly identified and taken into account before drawing any strong conclusions from lithium abundance data sets. The large size of our sample allows us to control these biases and to investigate any possible underlying connections between the presence of exoplanets and stellar surface lithium abundance.

Clearly, lithium abundances are very important for studies of stellar evolution and physics. By analyzing them, we hope to gain insights into not only stellar structure and evolution, but also the chemical evolution of lithium in the interstellar medium and, arguably, the impact of planet formation on stellar interior evolution.

\section{SAMPLE, OBSERVATIONAL DATA, AND LITHIUM ABUNDANCE ANALYSIS}

The data set analyzed in this work consists of about 700 high resolution ($R=\lambda/\Delta\lambda\gtrsim45,000$), high signal-to-noise ratio ($S/N\gtrsim100$) spectra of 671 nearby ($d < 200$ pc) FGK dwarf and subgiant stars taken from the RAL study. The high quality of these data allows a precise determination of lithium abundances even in stars with severe lithium depletion.

Most of the spectra were taken by us using the Tull coud\'e spectrograph on the 2.7\,m Harlan\,J.\,Smith Telescope at McDonald Observatory \citep{tull95}. We also used FGK stars' spectra from the online UVES-VLT library by \cite{bagnulo03} and the library of spectra of nearby stars --- the Spectroscopic Survey of Stars in the Solar Neighborhood, S$^4$N --- by \cite{allende04:s4n}, taken from both the Smith Telescope at McDonald Observatory and ESO's 1.5~m Telescope at La Silla. Some of RAL's spectra were not useful for lithium abundance determination due to relatively low S/N or incomplete wavelength coverage. Because of the difficulty of combining spectral data from sources with different spectral resolutions, when we had multiple spectra for a star, instead of co-adding the spectra, we independently analyzed them and averaged the abundances obtained.

We determined the lithium abundance using spectral line-profile fitting of the region near the lithium resonance doublet at 6708\,\AA, as in \cite{baumann10}. We used the spectrum synthesis code MOOG \cite[2010 version\footnote{http://www.as.utexas.edu/$\sim$chris/moog.html}]{sneden73}, which assumes local thermodynamic equilibrium (LTE), while the stellar parameters used are from RAL and were derived as described below.

The effective temperature ($\teff$) was obtained from the metallicity-dependent color-$\teff$ calibrations by \cite{casagrande10}. Most of our sample stars have ground-based measurements of $(B-V)$ and $(b-y)$ colors available in the General Catalog of Photometric Data \cite[GCPD,][]{mermilliod97}, the {\it Hipparcos} Catalog \citep{perryman97}, and the Geneva-Copenhagen Survey \citep{nordstrom04,casagrande11}. Stars fainter than about $V=6$ also have reliable infrared ($JHK$) photometry available from the Two-Micron All-Sky Survey \cite[2MASS][]{cutri03}. A small fraction of our sample stars also have $(RI)_\mathrm{C}$ photometry available in the GCPD. We used as many as possible of the color combinations that are available with these data to determine $\teff$, averaging different $\teff$ values derived from all available colors for each star (weighted by their errors, obtained by propagating the photometric uncertainties, the scatter of the color-$\teff$ calibrations, and considering also the error in $\feh$). The surface gravity ($\logg$) was derived from mass estimates based on theoretical isochrone analysis and trigonometric {\it Hipparcos} parallaxes. The isochrone set used is that by \cite{kim02}. Metallicities ([Fe/H])\footnote{In this work we use the standard definitions: $\mathrm{[X/Y]}=\log(N_\mathrm{X}/N_\mathrm{Y})-\log(N_\mathrm{X}/N_\mathrm{Y})_\odot$, and $A_\mathrm{X}=\log(N_\mathrm{X}/N_\mathrm{H})+12$, where $N_\mathrm{X}$ is the number density of element X.} were derived from equivalent width analysis of a large number of Fe\,\textsc{i} and Fe\,\textsc{ii} lines. Microturbulente velocities ($v_t$) were determined in the standard manner, i.e., by looking at the correlation between iron abundance and line strength of \fei\ lines in each spectrum, and iteratively increasing or decreasing the $v_t$ value until the slope of that correlation is close to zero.

The iron abundances inferred separately from the neutral and singly ionized species of iron agree satisfactorily. We find that their mean difference (\feii\---\fei) is $0.01\pm0.06$\,dex, implying that ionization balance is satisfied for the majority of the sample stars. The average internal errors in $\teff$, $\logg$, and $\feh$ are 53~K, 0.08~dex, and 0.05~dex, respectively. The mean value and error in [Fe/H] were obtained from the average and line-by-line scatter of \fei\ and \feii\ lines combined, respectively. Using all iron lines together for these calculations is justified by the fact that ionization balance is satisfied. The procedure described above was iterated several times because the stellar parameters are interdependent. We made sure that convergence to a final result was achieved for each star.

The model stellar atmospheres employed are from the MARCS grid with standard chemical composition, i.e., taking into account $\alpha$-element enhancement at low $\feh$.\footnote{http://marcs.astro.uu.se} Gaussian macroturbulent velocities were determined using the formulae by \cite{valenti05}, defined as the lower envelope of the macroturbulence as a function of $\teff$ derived by ignoring rotation. Rotational broadening was determined using line profile fitting of nearby Fe\,\textsc{i} lines, after applying instrumental and macroturbulent broadening.

Errors in lithium abundances due to uncertainties in stellar parameters were determined as follows. For each star, we generated a synthetic spectrum of the 6708\,\AA\ region using our derived stellar parameters and lithium abundance, without external broadening convolution. To calculate the error in lithium abundance due to the uncertainty in $\teff$, we increased the temperature of the model to $\teff+\Delta\teff$ where $\Delta\teff$ is the error in effective temperature according to RAL, and computed the lithium abundance necessary to match the first profile (that with our derived lithium abundance). We repeated the exercise for $\teff-\Delta\teff$. The average of the differences between these new lithium abundances and the original one was adopted as the error due to the $\teff$ uncertainty. A similar procedure was used to measure the smaller errors due to $\logg$ and [Fe/H] uncertainties. The three errors were then added in quadrature. This is a reasonable approximation considering that the atmospheric parameters were derived by RAL using nearly independent techniques.\footnote{The stellar parameters derived by RAL are not fully independent, but their correlation is weak and their impact on the errors is smaller than the internal uncertainties. For example, for a Sun-like star, a change of 50\,K in $\teff$ corresponds to 0.03\,dex for $\feh$ (if inferred from \fei\ lines, but smaller if derived from \feii). Propagating this $\feh$ shift to re-determine $\teff$ leads to a change of 8\,K, which is a small fraction of the original $\teff$ error.}

For the majority of our sample stars, the error in lithium abundance due to the synthesis fitting procedure itself (i.e., the error due to observational noise) was found to be negligible compared to the error due to stellar atmosphere uncertainties. A few noisy spectra were the exception. They were assigned a conservative error of 0.15\,dex. This is larger than any error determined for our sample stars for which lithium abundances were measured. In cases where the lithium doublet was not detected, we were still able to calculate an upper limit to the lithium abundance based on the stellar parameters of the star and the local signal-to-noise ratio of the spectra (these values correspond to 2\,$\sigma$ detection limits). Our lithium abundance results are given in Table \ref{tbl_ours}.\footnote{RAL's data set also includes 9 solar spectra, either from (reflected sunlight) asteroid or skylight observations. The average LTE lithium abundance for these spectra is 1.02~dex, with a standard error of 0.02 dex. This value is in good agreement with other recent determinations of the solar lithium abundance \cite[e.g.,][]{asplund09:review}.}

\begin{deluxetable}{crrrr}
\tablecolumns{5}
\tablewidth{0pc}
\tablecaption{Stellar Parameters and Lithium Abundances Determined in This Work.\tablenotemark{1}}
\tabletypesize{\footnotesize}
\tablehead{\colhead{HIP} & \colhead{$\teff$ (K)} & \colhead{$\logg$ [cgs]} & \colhead{[Fe/H]} & \colhead{$\ali$}}
\startdata
171 & $5510\pm66$ & $4.46\pm0.01$ & $-0.76\pm0.06$ & $<0.35$ \\
348 & $5746\pm55$ & $4.38\pm0.06$ & $-0.19\pm0.04$ & $0.82\pm0.04$ \\
394 & $5635\pm37$ & $3.78\pm0.07$ & $-0.48\pm0.04$ & $1.98\pm0.04$ \\
475 & $5836\pm72$ & $4.35\pm0.05$ & $-0.34\pm0.06$ & $1.63\pm0.05$ \\
493 & $5960\pm44$ & $4.41\pm0.03$ & $-0.20\pm0.04$ & $2.32\pm0.04$ \\
522 & $6251\pm44$ & $4.21\pm0.02$ & $0.05\pm0.05$ & $2.84\pm0.04$ \\
530 & $5866\pm40$ & $3.90\pm0.05$ & $-0.01\pm0.05$ & $0.70\pm0.03$ \\
544 & $5458\pm40$ & $4.52\pm0.02$ & $0.14\pm0.06$ & $2.39\pm0.04$ \\
656 & $5805\pm39$ & $3.82\pm0.06$ & $-0.24\pm0.06$ & $2.53\pm0.04$ \\
768 & $5681\pm31$ & $3.99\pm0.04$ & $-0.01\pm0.06$ & $1.46\pm0.02$ \\

\vdots & \vdots & \vdots & \vdots & \vdots 
\enddata
\tablenotetext{1}{Table~\ref{tbl_ours} is published in its entirety in the electronic edition of the Astrophysical Journal. A portion is shown here for guidance regarding its form and content.}
\label{tbl_ours}
\end{deluxetable}

\section{A CATALOG OF LITHIUM ABUNDANCES}

\begin{deluxetable*}{lrrrrrr}
\tablecolumns{6}
\tablewidth{0pc}
\tablecaption{Offsets in Stellar Parameters and Lithium Abundances Relative to This Work.\tablenotemark{1}}
\tabletypesize{\footnotesize}
\tablenotetext{1}{The errors quoted for the $\Delta$ values correspond to the 1\,$\sigma$ standard deviation of the differences.}
\tablehead{\colhead{Sample\tablenotemark{a}} & \colhead{$\Delta\teff$ (K)} & \colhead{$\Delta\logg$} & \colhead{$\Delta\feh$} & \colhead{$\Delta\ali$} & \colhead{$N$\tablenotemark{b}}  & \colhead{$N_\mathrm{tot}$\tablenotemark{c}}}
\startdata
 B10 & $ 40\pm48$ & $ 0.05\pm0.06$ & $ 0.03\pm0.03$ & $ 0.07\pm0.09$ &  69 & 117 \\
 G10 & $ -4\pm68$ & $-0.02\pm0.04$ & $ 0.04\pm0.04$ & $-0.13\pm0.12$ &  44 & 152 \\
 I09 & $ 26\pm32$ & $ 0.05\pm0.05$ & $ 0.02\pm0.02$ & $ 0.06\pm0.04$ &  11 &  80 \\
LH06 & $ 50\pm85$ & $ 0.04\pm0.19$ & $ 0.01\pm0.06$ & $ 0.01\pm0.11$ &  76 & 194 \\
 T10 & $-12\pm54$ & $-0.03\pm0.09$ & $ 0.03\pm0.04$ & $-0.03\pm0.10$ &  26 & 117 \\
Gh10 & $ 30\pm71$ & $-0.06\pm0.14$ & $-0.00\pm0.06$ & $ 0.01\pm0.08$ &  38 & 259 \\
LR04 & $-83\pm63$ & $-0.01\pm0.11$ & $-0.02\pm0.05$ & $-0.07\pm0.08$ & 212 & 449 \\

\enddata
\tablenotetext{a}{Sample: B10 = \cite{baumann10}, G10 = \cite{gonzalez10:lithium}, I09 = \cite{israelian09}, LH06 = \cite{luck06}, T10 = \cite{takeda10}, Gh10 = \cite{ghezzi10}, LR04 = \cite{lambert04}.}
\tablenotetext{b}{$N$ is the number of stars in common with this work.}
\tablenotetext{c}{$N_\mathrm{tot}$ is the number of stars in our final catalog taken from each study.}
\label{t:offsets}
\end{deluxetable*}

We supplemented our measurements with lithium abundances from \cite{lambert04,luck06,israelian09,baumann10,ghezzi10,gonzalez10:lithium}; and \cite{takeda10}. \cite{lambert04} included a compilation of data from \cite{balachandran90,chen01}; and \cite{reddy03}, which they ensured to be consistent with regards to the stellar parameters and lithium abundances. All of these works used high quality data and their results are, internally, analyzed in a homogeneous fashion. These previously published data were normalized to ours by computing and subtracting the mean offsets in $\teff$, $\logg$, [Fe/H], and lithium abundance, to compensate for systematic differences in stellar parameters which may affect our mass and age determinations. Stars in common between this work and the previously published studies were used to calculate these offsets, which are given in Table~\ref{t:offsets}.

Instead of deriving lithium abundance offsets based on the mean differences in stellar parameters, we used the mean offsets in lithium abundances themselves to put all lithium abundance measurements into the same scale. This is because the lithium abundances depend not only on the stellar parameters adopted by each group, but also on the atomic data employed in the spectral line synthesis of the 6708\,\AA\ lithium doublet. Upper limits to lithium abundances were excluded from the computation of the offsets. \cite{lambert04} and \cite{takeda10} applied non-LTE corrections to their lithium abundances, which we undid in order to prevent different non-LTE prescriptions from systematically biasing our data, i.e., we recovered their LTE lithium abundances.

When a star was available in more than one source, we calculated a weighted average for each of the parameters and lithium abundance. We adopted as weights the inverse variances given in each of the sources. The errors quoted by \cite{israelian09} are very small compared to those from other sources. Such small errors are probably reasonable for internal analysis of their sample, but systematic errors will dominate when comparing to other works. In order to avoid giving too much weight to the \cite{israelian09} data (no source should be given very high preference in this type of compilation), we increased their error bars by a factor of 3 in $\teff$ and $\logg$ and a factor of 4 in [Fe/H] (retaining the relative size of the errors equal, therefore giving more weight to stars which were analyzed using higher quality data, or stars that are more similar to the Sun, where systematic errors are smaller). These multiplicative factors were estimated from the comparison of our data to theirs. The star-to-star scatter in the differences in the parameters for stars in common could be made compatible with the quoted errors if increased by those factors.

For \cite{lambert04}, \cite{luck06}, and \cite{takeda10}, we adopted errors of 50\,K, 0.07 ($\logg$), and 0.05 ([Fe/H]) for all stars. Although more realistic errors, generally including systematics, of about 100\,K, 0.1\,dex, and 0.1\,dex, respectively, are typically quoted in the literature, we adopted these error bars to give a reasonable weight to these stellar parameter and lithium abundance measurements when computing average values within our compilation.

Our procedure of increasing errors as described in the previous paragraphs may appear somewhat arbitrary, although we have given a quantitative justification for it. Nevertheless, we note that not weighting the averages gives very similar results. Indeed, for the 475 stars which have data from more than one source, the difference in $\teff$ between the weighted and non-weighted values is $0\pm12$\,K, that for $\logg$ is $0.1\pm0.2$, and for $\feh$ we find $-0.001\pm0.011$. These small mean differences have only a marginal effect on our subsequent analysis.

In our compilation, when lithium upper limits were available from more than one source, we adopted the lowest upper limit found, which typically corresponds to the case where the highest quality data were used.

To correct for non-LTE effects on the 6708\,\AA\ lithium doublet formation and abundance derivation, we used the non-LTE tables and interpolation code by \cite{lind09}. For a number of stars, the parameters (especially microturbulent velocity) were outside of the range covered by these tables. For these stars, we used the nearest node in the stellar parameter grid of the non-LTE tables to determine the size of the correction. These extrapolations are not risky because of the asymptotic nature of the non-LTE corrections in these extremes of the stellar parameter grid used by \cite{lind09}. For most of our sample stars, the non-LTE corrections are of order +0.02\,dex, and more important ($\simeq-0.05$\,dex) only for stars with very high lithium abundance ($\ali\gtrsim2.5$).

\begin{deluxetable*}{crrrrrrrrr}
\tablecolumns{5}
\tablewidth{0pc}
\tablecaption{Catalog of Stellar Parameters and Lithium Abundances.\tablenotemark{1}}
\tabletypesize{\footnotesize}
\tablehead{\colhead{HIP} & \colhead{$\teff$ (K)} & \colhead{$\logg$} & \colhead{[Fe/H]} & \colhead{$\ali$} & \colhead{Mass $M_\odot$} & \colhead{Age (Gyr)} & \colhead{Pop.\tablenotemark{a}} & \colhead{Planets} & \colhead{Source\tablenotemark{b}}}
\startdata
171 & $5449\pm77$ & $4.46\pm0.01$ & $-0.84\pm0.08$ & $<-0.51$ & $0.76^{+0.01}_{-0.01}$ & $6.3\pm1.4\mathrm{\ [rot]}$ & \nodata & \nodata & TW+LH06 \\
348 & $5740\pm50$ & $4.38\pm0.06$ & $-0.17\pm0.04$ & $0.85\pm0.07$ & $0.94^{+0.03}_{-0.02}$ & $7.66^{+2.35}_{-2.33}\mathrm{\ [iso]}$ & thin & no & TW+B10 \\
394 & $5636\pm50$ & $3.76\pm0.07$ & $-0.49\pm0.04$ & $1.98\pm0.02$ & $1.10^{+0.15}_{-0.05}$ & $5.00^{+2.29}_{-1.03}\mathrm{\ [iso]}$ & \nodata & no & TW+LR04 \\
475 & $5836\pm72$ & $4.36\pm0.05$ & $-0.34\pm0.06$ & $1.63\pm0.05$ & $0.92^{+0.03}_{-0.03}$ & $8.75^{+2.72}_{-1.65}\mathrm{\ [iso]}$ & thin & no & TW \\
493 & $5943\pm50$ & $4.38\pm0.03$ & $-0.23\pm0.04$ & $2.31\pm0.02$ & $0.99^{+0.03}_{-0.02}$ & $5.77^{+1.13}_{-2.00}\mathrm{\ [iso]}$ & thin & \nodata & TW+LR04 \\
522 & $6266\pm50$ & $4.21\pm0.02$ & $0.04\pm0.04$ & $2.84\pm0.02$ & $1.24^{+0.01}_{-0.01}$ & $2.90^{+0.42}_{-0.24}\mathrm{\ [iso]}$ & thin & yes & TW+Gh10 \\
530 & $5868\pm50$ & $3.89\pm0.05$ & $-0.05\pm0.04$ & $0.70\pm0.03$ & $1.20^{+0.08}_{-0.03}$ & $4.90^{+0.75}_{-0.48}\mathrm{\ [iso]}$ & thin & no & TW+LR04 \\
544 & $5465\pm50$ & $4.53\pm0.03$ & $0.14\pm0.04$ & $2.38\pm0.04$ & $0.96^{+0.02}_{-0.03}$ & $0.2\pm0.1\mathrm{\ [rot]}$ & thin & no & TW+LH06 \\
656 & $5819\pm50$ & $3.82\pm0.06$ & $-0.25\pm0.04$ & $2.53\pm0.02$ & $1.25^{+0.09}_{-0.07}$ & $3.89^{+1.53}_{-0.33}\mathrm{\ [iso]}$ & \nodata & \nodata & TW+LR04 \\
699 & $6179\pm63$ & $4.13\pm0.04$ & $-0.30\pm0.05$ & $2.33\pm0.10$ & $1.11^{+0.04}_{-0.01}$ & $5.50^{+0.29}_{-1.07}\mathrm{\ [iso]}$ & thin & no & LR04 \\

\vdots & \vdots & \vdots & \vdots & \vdots & \vdots & \vdots & \vdots & \vdots & \vdots
\enddata
\tablenotetext{1}{Table~\ref{tbl_all} is published in its entirety in the electronic edition of the Astrophysical Journal. A portion is shown here for guidance regarding its form and content.}
\tablenotetext{a}{Population: thin-disk, thick-disk, or halo, according to the star's kinematics. If not available, the kinematic criterion adopted in this work implies an ambiguous or unreliable association.}
\tablenotetext{b}{Source: TW = This Work, B10 = \cite{baumann10}, G10 = \cite{gonzalez10:lithium}, I09 = \cite{israelian09}, LH06 = \cite{luck06}, T10 = \cite{takeda10}, Gh10 = \cite{ghezzi10}, LR04 = \cite{lambert04}.}
\label{tbl_all}
\end{deluxetable*}

Stellar masses and ages were derived as in RAL. The method is also described in \cite{baumann10,melendez12}; and \cite{chaname12}. In summary, each star is placed on the HR diagram using its $\teff$ and absolute magnitude. The latter was computed using parallaxes from the new reduction of {\it Hipparcos} data \citep{vanleeuwen07}. All theoretical isochrone points from the Yonsei-Yale grid \cite[e.g.,][]{yi01,kim02} within a radius of $3\,\sigma$ from the observed parameters were then used to compute mass and age probability distribution functions, from which the most probable parameters and $1\,\sigma$-like errors were computed. The same procedure was used to determine trigonometric surface gravities. These $\logg$ values are internally consistent with the $\teff$ and $\feh$ parameters adopted for each star in this work, and we prefer them to those derived as the average of published values.

Isochrone ages are unreliable for young stars. These objects, however, tend to be active and their rotation periods can be measured from the modulation due to starspots on photometric light curves and/or chromospheric emission data. Rotation periods can then be used to estimate stellar ages more precisely than using isochrones, particularly for young stars, because stars with convective envelopes slow down in a predictable way as they age \cite[e.g.,][]{barnes07}. Therefore, if available, we used measured rotation periods ($P_\mathrm{rot}$) and the age-$P_\mathrm{rot}$ relation by \cite{barnes07} to estimate the ages of stars. Measurements of rotation period adopted in this work are from \cite{gaidos00,strassmeier00}; and \cite{pizzolato03}.

All the stars in our sample have kinematical data from RAL, namely radial velocities and Galactic space velocities inferred from them and the stars' parallaxes and proper motions. From these data we compute a kinematical probability that a star belongs to the thin disk ($P_1$), thick disk ($P_2$), and halo ($P_3$). The formulae used to calculate these probabilities are given in, for example, \cite{mishenina04} and \cite{ramirez07}. To assign a given star to one of these groups, we adopt a criterion stronger than that employed by \cite{ramirez07}, which is $P_1>0.7$ for thin-disk stars and $P_2>0.7$ for thick-disk stars. In addition to those conditions, here we require $P_1/P_2>10$ for thin-disk stars and $P_2/P_1>10$ for thick-disk members, as suggested by \cite{bensby05}. For halo stars we adopt $P_3>0.5$. With these constraints we find in our sample 898 thin-disk stars, 144 thick-disk stars, and 43 halo members. The rest of our targets (296 stars) have ambiguous or uncertain population membership.

The stars in our catalog are also flagged as known planet-hosts (Planets=``yes''), non planet-hosts (Planets=``no''), and stars whose planetary status is unknown (Planets=``\nodata''). The non planet-host stars have had their radial velocities monitored but no planets have been detected around them yet. We used tables by \cite{fischer05,sousa08,israelian09,ghezzi10,brugamyer11}, and \cite{wright11} to sort the stars into these categories. We found that 165 stars in our catalog are known to have planets while 360 stars are, as of this writing, deemed to have no sub-stellar mass companions.

Our catalog of stellar parameters, lithium abundances, and derived masses and ages is given in Table~\ref{tbl_all}. The errors for the $\teff$ and $\feh$ values given in Table~\ref{tbl_all} correspond to the sample variance in cases where data from more than one source were available. However, we adopted minimum errors of 50\,K in $\teff$ and 0.04\,dex in $\feh$ to prevent very small unrealistic errors arising from coincidental agreement of the statistically few sources available. The mean errors in $\teff$, $\logg$, and $\feh$ given in this Table are 59\,K, 0.08\,dex, and 0.05\,dex, respectively, while that for the lithium abundances (excluding the upper limits) is 0.08\,dex. In Table~\ref{tbl_all} we also provide population and planet flags, as described in the previous paragraphs, and the source(s) from which the data were taken.

\section{DISCUSSION}

\subsection{The Lithium Desert} \label{s:desert}

\begin{figure}
\includegraphics[bb=90 375 560 740,width=8.9cm]
{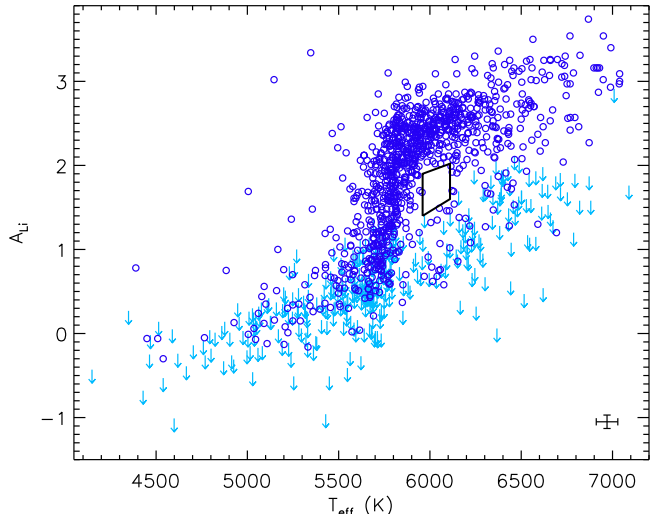}
\caption{Lithium abundance as a function of effective temperature for all stars in our catalog. Downward arrows represent upper limits to the lithium abundance. The polygon bounds the approximate extent of the lithium desert and it is drawn here and in subsequent figures only to guide the eye. An error bar representing the average internal errors is shown at the bottom right.}
\label{f:li_teff}
\end{figure}

Figure~\ref{f:li_teff} shows our lithium abundances and effective temperatures. Overall, lithium is less abundant in cooler, less massive stars with larger convective envelopes. The two stars with $\teff<5500$\,K and high lithium abundance ($\ali>2.8$) are HIP\,13402 and HIP\,30034. Both stars are known to be very young, which means that their surface lithium abundances have not yet been depleted by internal processes. HIP\,13402 has an age of a few million years \cite[e.g.,][]{cayrel89} while HIP\,30034 is a pre-main sequence star \cite[e.g.,][]{scholz07}.

One of the most striking features of our lithium abundance compilation, first suggested by \cite{chen01}, is seen clearly in Figure~\ref{f:li_teff}. Between $\teff\simeq5950$\,K and 6100\,K, the stars appear to separate into two groups of high and low lithium abundance, creating a ``lithium desert,'' a region of the $\teff$-$\ali$ plane that seems to not contain any stars even though it is surrounded by them. This region corresponds to the area inside the polygon plotted in Figure~\ref{f:li_teff}. The lithium desert likely extends to slightly cooler and warmer effective temperatures, but its boundaries are blurred by our $\teff$ uncertainties. The extent of the lithium desert in both $\teff$ and $\ali$, and the abundance of stars immediately outside of its boundaries make it highly unlikely that this feature is a product of our sample selection and/or observational errors.

\begin{figure*}
\includegraphics[bb=80 365 960 1085,width=18.3cm]{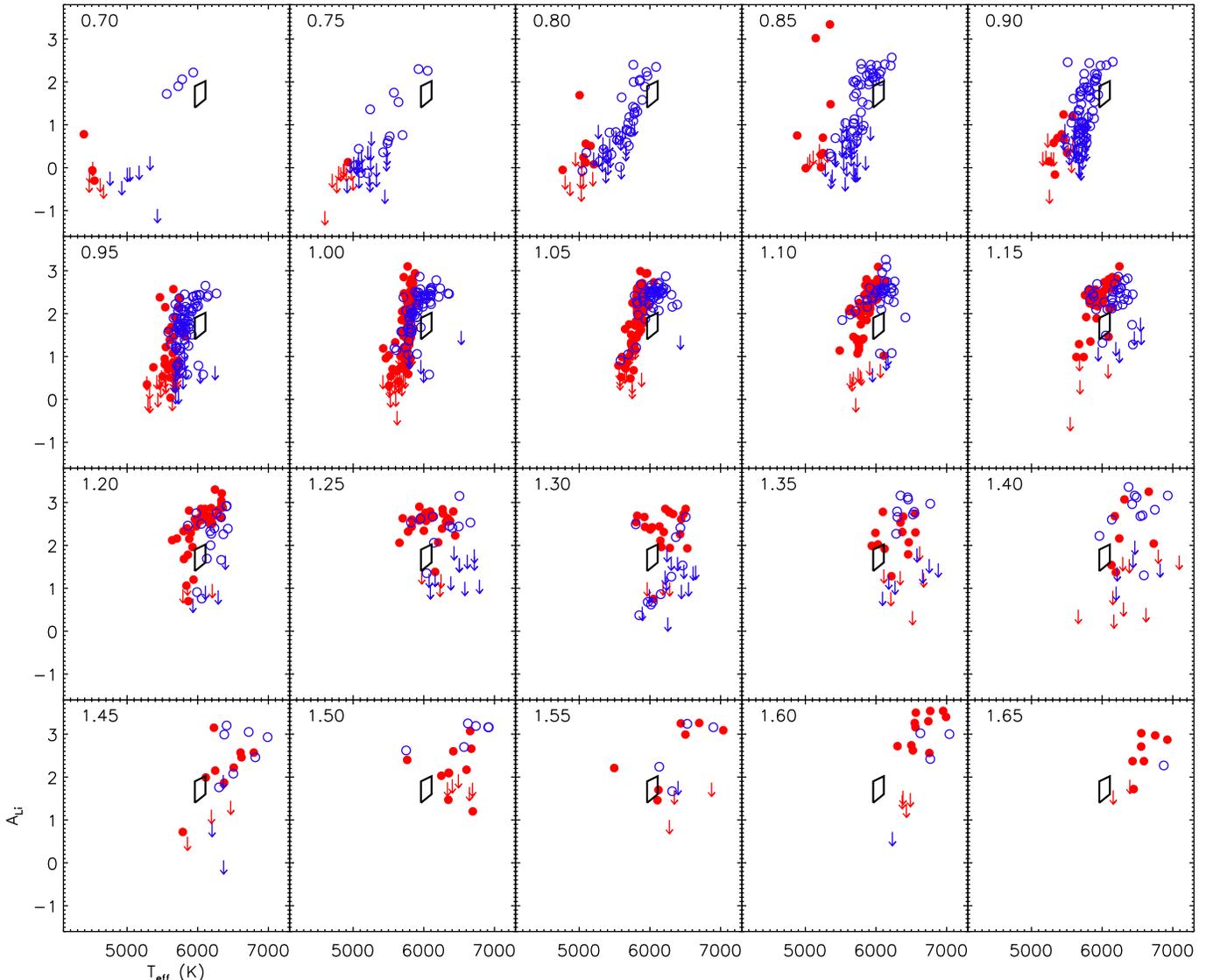}
\caption{Lithium abundance as a function of effective temperature for stars of a given mass and metallicity. In each panel, stars have a mass within $0.05\,M_\odot$ from the number shown on the top left. Filled (open) circles correspond to stars more metal-rich (metal-poor) than $\feh=-0.1$. Downward arrows represent upper limits to the lithium abundance. The polygon shows the location of the lithium desert.}
\label{f:li_teff_dissect}
\end{figure*}

\begin{figure*}
\includegraphics[bb=80 365 960 1085,width=18.3cm]{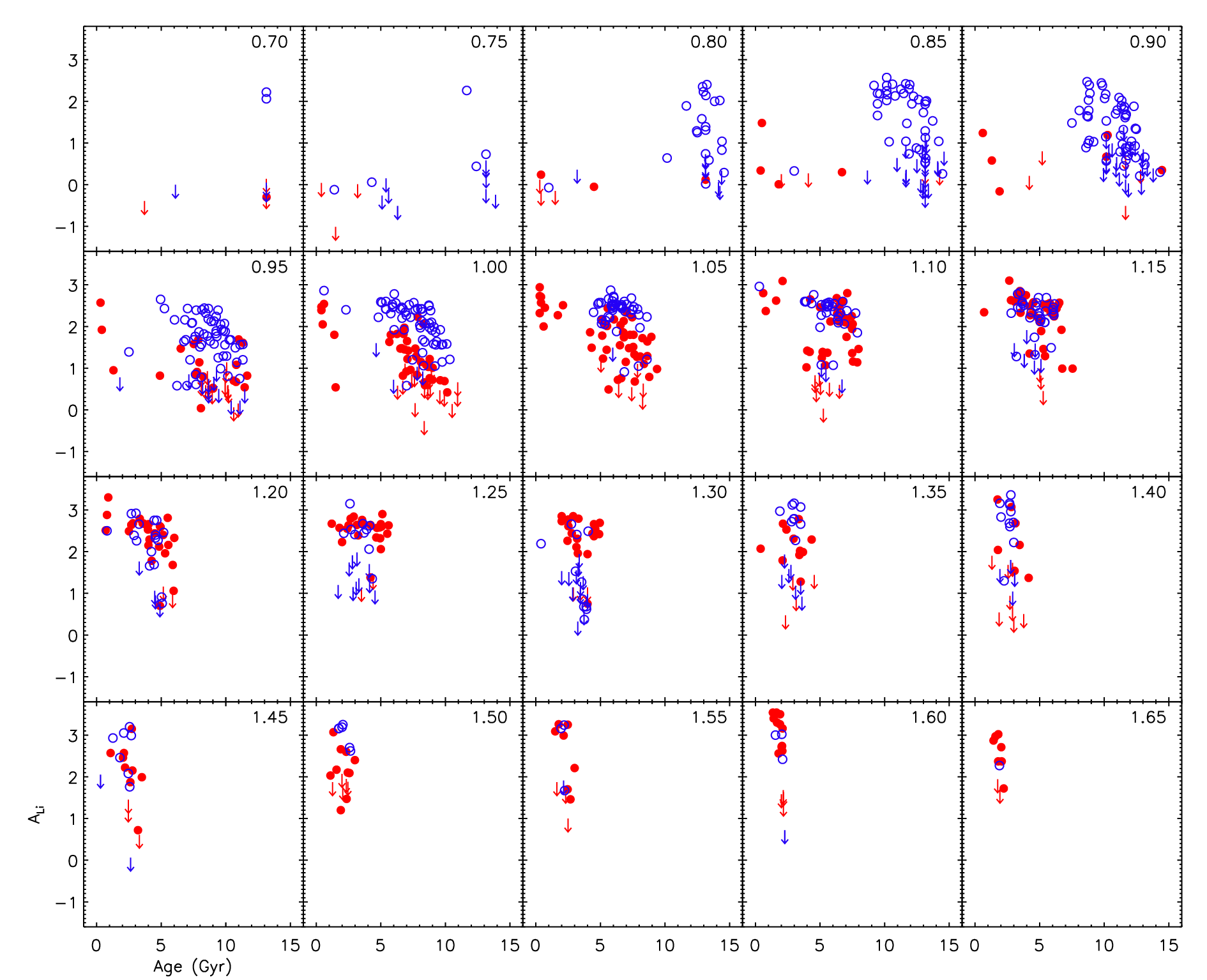}
\caption{Lithium abundance as a function of age for stars of a given mass and metallicity. In each panel, stars have a mass within $0.05\,M_\odot$ from the number shown on the top right. Filled (open) circles correspond to stars more metal-rich (metal-poor) than $\feh=-0.1$. Downward arrows represent upper limits to the lithium abundance.}
\label{f:li_age_dissect}
\end{figure*}

Since many, but not all, of the stars in the lower part of the lithium desert have only upper limits for their lithium abundances, i.e., their $\ali$ values are based on the non-detection of the 6708\,\AA\ lithium doublet, the positive slope of the $\ali$ versus $\teff$ relation for those stars simply reflects the detection limit of the lithium feature for spectra with signal-to-noise ratio typical of that of our sample stars and those taken from the literature. The maximum $\ali$ value possible for a given star with no lithium doublet detection depends on $\teff$ because the minimum line equivalent width detectable, which is a constant assuming all spectra have the same signal-to-noise ratio,  requires a higher lithium abundance for higher $\teff$ values. This probably explains the shape of the lower envelope of the $\ali$ versus $\teff$ relation plotted in Figure~\ref{f:li_teff}, not only around the lithium desert, but also over the entire $\teff$ range covered by the stars in our compilation.

In order to investigate the nature of the lithium desert, as well as other patterns of lithium depletion across the stellar parameter space covered by our sample, in Figure~\ref{f:li_teff_dissect} we show $\ali$ versus $\teff$ trends for stars of a given mass. Moreover, within each mass group, we divide the stars into two groups of metallicity, a ``metal-rich'' group of $\feh>-0.1$ (filled circles) and a ``metal-poor'' group of $\feh<-0.1$ (open circles).

As expected, there is a strong correlation between $\teff$ and mass -- all our sample stars are on the main sequence or are slightly evolved. The well-known effect that at any given mass lower metallicity stars have higher surface temperatures is also obvious in Figure~\ref{f:li_teff_dissect}.

Stars with masses between 0.70 and $1.35\,M_\odot$ can be found on the high lithium abundance side of the lithium desert, but, with very few exceptions, stars on the low lithium abundance side of the desert appear to have $M>1.10\,M_\odot$. Moreover, at any given mass, stars in the lower part of the desert tend to be more metal poor than those with higher lithium abundances, an effect that is clearly seen in the case of $1.30\,M_\odot$ stars. Due to their warmer effective temperatures, most stars with $M>1.4\,M_\odot$ are found beyond the lithium desert.

Lithium abundance versus age trends for stars of a given mass (and metallicity) are shown in Figure~\ref{f:li_age_dissect}. Only stars with reliable ages (age/error\,$>3$) are plotted there. For stars more massive than $M=0.8\,M_\odot$ there is some evidence of main-sequence lithium depletion, but the time-scales seem to depend not only on mass but also on the stellar metallicity. For $M<1.1\,M_\odot$, metal-rich stars seem to deplete lithium quicker than metal-poor stars, a feature that is particularly dramatic in the $1.00\pm0.05\,M_\odot$ group. Interestingly, this pattern appears to invert at higher masses, although the trends become less clearly defined. This picture is further complicated by the fact that the metallicity dependence of main-sequence lithium depletion seems to disappear for stars with $M>1.35\,M_\odot$.

Figures~\ref{f:li_teff_dissect} and \ref{f:li_age_dissect} show that, in addition to the stellar mass, metallicity is a key parameter to understand surface lithium depletion. This is not entirely surprising, since the properties of the convective zone, in particular the temperature at the base, have a high sensitivity to both parameters \cite[e.g.,][]{pinsonneault01,theado12}.

Although the morphology of the lithium desert is accentuated by the combination of low and high mass stars in Figure~\ref{f:li_teff}, it can be still detected in plots showing stars within a narrow mass range (Figure~\ref{f:li_teff_dissect}). From $M=1.10$ to $1.30\,M_\odot$, stars within a mass range of $0.05\,M_\odot$ of a given mass are found both below and above the desert. Figure~\ref{f:li_age_dissect} does not show a sudden drop in the surface lithium abundance at a given age for stars in that mass range, although errors in our age determination might be preventing us from finding such trend. In these groups, however, and in particular in the $1.10\,M_\odot$ group, stars appear to separate into high and low lithium abundance subgroups in a discontinuous manner. It is also clear that there are no young stars ($\mathrm{age}\lesssim2$\,Gyr) in the lower part of the lithium desert while those that appear relatively young ($\mathrm{age}\simeq3$\,Gyr) are also the most massive ones. Since the main-sequence lifetime of a $1.1\,M_\odot$ star is about 7.5\,Gyr, the mechanism responsible for the big drop in surface lithium abundance, and hence the lithium desert, must act while the star is on the main sequence or as a subgiant star, and most likely not in the first 2\,Gyr. We cannot rule out the possibility that this process is very short lived, as the dramatic, complete lack of stars in the desert suggests.

\cite{chen01} argued that the stars on the low lithium abundance side of the desert are evolved lithium dip stars. They found a very strong and tight correlation between mass and metallicity for these low lithium abundance stars, which they used to support their idea, because the mass of the lithium dip stars seems to correlate with the metallicity of the cluster \cite[e.g.,][]{balachandran95}. However, using their data (i.e., their masses and metallicities), we find a similarly tight correlation for their high lithium abundance stars, implying that the mass-$\feh$ relation is inherent to the data set and sample selection, rather than a property of groups of stars with different lithium abundance patterns.

The mass-$\feh$ correlation described above could also be due to systematic biases in the determination of masses. Using our data, we also find a correlation between mass and $\feh$ for the stars on the low lithium abundance side of the desert, but the dispersion is larger than that found by \cite{chen01}. In any case, we find a similar behavior for the stars with high lithium abundance. Another possible reason for this correlation could be simply that the $\teff$ range of the desert is so narrow that the metallicity effect on the location of the main-sequence, which becomes redder as $\feh$ increases, essentially projects into a mass dependence. Thus, our analysis suggests that the lithium desert cannot be explained only with evolved lithium dip stars. A possibly short-lived mechanism operating during the main sequence or subgiant phase of stellar evolution is likely responsible for this feature.

\subsection{Galactic Chemical Evolution of Lithium}

\subsubsection{The Lithium Problem}

A conventional procedure for probing the Galactic chemical evolution (GCE) of elements is to establish the run of [X/H] or [X/Fe] with $\feh$ from spectroscopic analyses of samples of stars such as those in the solar neighborhood. In the case of lithium, since surface depletions are common and may be severe, the custom has been to plot $\ali$ versus [Fe/H]  and to adopt the upper envelope of the distribution as defining the growth of lithium abundance with the iron abundance. Then, this upper envelope is confronted with theories of lithium nucleosynthesis and evolution of the Galaxy \citep[e.g.,][]{ryan01,travaglio01,alibes02}.

Our catalog of lithium abundances, if analyzed in the above fashion, would confirm previous attempts to map the GCE of lithium. Often overlooked in these attempted mappings is the question of lithium depletion in those stars defining the upper envelope. This issue is most obviously a concern at low metallicities where the upper envelope is defined by the so-called Spite plateau \citep{spite82}, for which $\ali\simeq2.2$. Now that observations from the Wilkinson Microwave Anisotropy Probe (\textit{WMAP}) have unequivocally defined the baryon-to-photon ratio, standard Big Bang nucleosynthesis predicts a primordial lithium abundance of $\ali\simeq2.7$ \cite[e.g.,][]{richard05,steigman10}, and this has led to the so-called lithium problem \cite[e.g.,][]{fields11}, for which two broad solutions are in contention: 1) have the stars on the Spite plateau depleted their atmospheric lithium abundance by about 0.5 dex? or 2) do the standard Big Bang nucleosynthesis calculations need reconsideration on account of revisions to the standard models of particle physics and/or cosmology? Given that the present (meteoritic) lithium abundance is $\ali\simeq3.3$, an estimate free from uncertainty about stellar depletion effects, a change in the pre-Galactic abundance from $\ali\simeq2.2$ to 2.7 profoundly reduces the inferred growth rate of lithium in the Galaxy.

Depletion in metal-poor low mass stars would imply depletion may also occur in the less metal-poor and somewhat higher mass stars which define the  start of the upper envelope's transition from the Spite plateau to higher lithium abundances in disk stars. Mass estimates for stars in our catalog show that the maximum mass of stars populating the $\ali$ versus $\feh$ upper envelope decreases with decreasing [Fe/H] (see also \citealt{chen01} and \citealt{lambert04}), from about $1.6\,M_\odot$ at $\feh=0$ to $1.0\,M_\odot$ at $\feh=-1$. At the lower masses, for example $M<1.3M_\odot$, lithium is depleted in main-sequence stars with depletion increasing with decreasing mass and increasing age, as clearly shown by the lithium abundance reported for open clusters --- see, for example, \cite{sestito05}, who, however, argue that depletion becomes ineffective for stars older than about 1--2 Gyr. Thus, the $\ali$ versus $\feh$ upper envelope most likely does not represent the growth of Galactic lithium abundance with [Fe/H]. The observed lithium abundances should be revised upwards by an amount that increases with decreasing [Fe/H].

\subsubsection{Lithium in the Thin and Thick Disks}

It is well known that, at least in the solar neighborhood, thin- and thick-disk stars have different chemical composition \cite[e.g.,][]{bensby05,reddy06}. Thus, one expects that thin and thick disks may have different patterns for their lithium versus iron abundance evolution. Yet, to the best of our knowledge, no attempts have been made to find these different patterns. Our sample is of sufficient size and diversity that an attempt may now be made. Unfortunately, unlike the elements which have been used to define
the thin/thick disk differences in composition, lithium is subject, as noted above, to internal depletion; different $\ali$ versus [Fe/H] patterns for thin and thick disk may, therefore, reflect depletion dependencies rather than different histories of lithium nucleosynthesis.

\begin{figure}
\includegraphics[bb=90 370 450 745,width=8.9cm]
{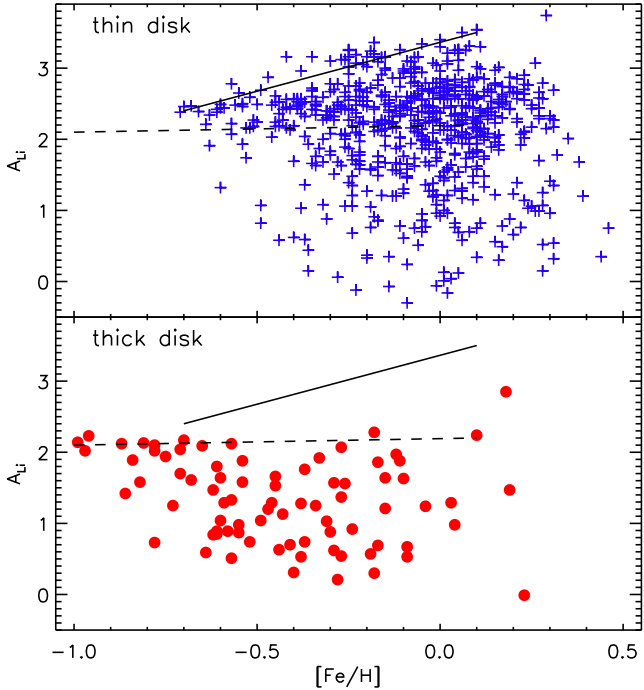}
\caption{Lithium abundance as a function of $\feh$ for thin-disk (upper panel) and thick-disk (lower panel) stars. Lines are drawn ``by hand'' to make the upper envelope differences clear. Solid (dashed) lines trace the upper envelope of the thin-disk (thick-disk) star data. Upper limits to lithium abundances based on the non detection of the lithium 6708\,\AA\ doublet are not shown.}
\label{f:gce1}
\end{figure}

Lithium abundance against metallicity is plotted in Figure~\ref{f:gce1} separately for thin- and thick-disk stars.\footnote{The thick-disk star with the highest lithium abundance in the lower panel of Figure~\ref{f:gce1} is HIP\,10856 ($\ali=2.85\pm0.04$, $\feh=+0.18\pm0.06$). This object has a high radial velocity and a high proper motion, which make its Galactic space velocities typical of a thick-disk star. Its atmospheric parameters, however, suggest that this object is younger than about 4\,Gyr. We note also that this star has a level of chromospheric emission above that of most stars of similar color \citep{isaacson10}, confirming its relatively young age. Moreover, its derived mass ($1.18\pm0.03\,M_\odot$) implies a lifetime of about 6\,Gyr. Thus, instead of being a true thick-disk member, it seems more likely that the orbit of this star has been excited by a collision. If that were true, the highest $\ali$ point in the bottom panel of Figure~\ref{f:gce1} should be in the top panel instead, where it would fit the overall trend perfectly.\label{tkta}} Our data clearly indicate different trends for the two disks. The full sample of thin-disk stars (Figure~\ref{f:gce1}, upper panel) shows an apparent enrichment with increasing metallicity, contrary to the case of thick-disk stars (Figure~\ref{f:gce1}, lower panel) which appear to have a nearly constant maximum lithium abundance. Interestingly, this maximum lithium abundance is similar to that of the Spite plateau exhibited by very metal-poor stars.

However, the thin/thick disk difference shown in Figure~\ref{f:gce1} may be influenced (likely dominated) by systematic differences in mass, metallicity, and age and, therefore, lithium depletions between the two samples.

\begin{figure}
\includegraphics[bb=80 368 450 1028,width=8.9cm]
{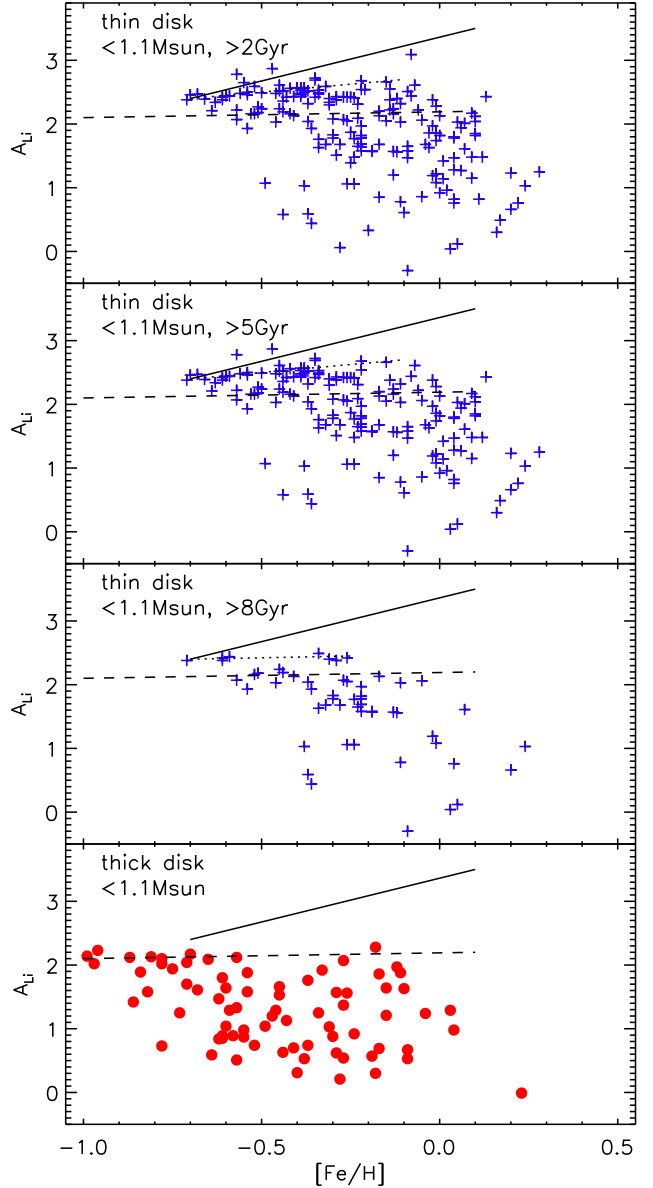}
\caption{Lithium abundance as a function of $\feh$ for thin-disk (upper three panels) and thick-disk (lower panel) stars of mass $M<1.1\,M_\odot$. In the upper three panels, only thin-disk stars older than 2, 5, and 8\,Gyr, respectively, are shown. Lines are drawn ``by hand'' to make the upper envelope differences clear. Solid (dashed) lines trace the upper envelope of the thin-disk (thick-disk) star data. The dotted lines trace the upper envelope of thin-disk stars older than 2, 5, and 8\,Gyr, respectively. Upper limits to lithium abundances based on the non detection of the lithium 6708\,\AA\ doublet are not shown.}
\label{f:gce2}
\end{figure}

\begin{figure}
\includegraphics[bb=90 375 450 600,width=8.9cm]
{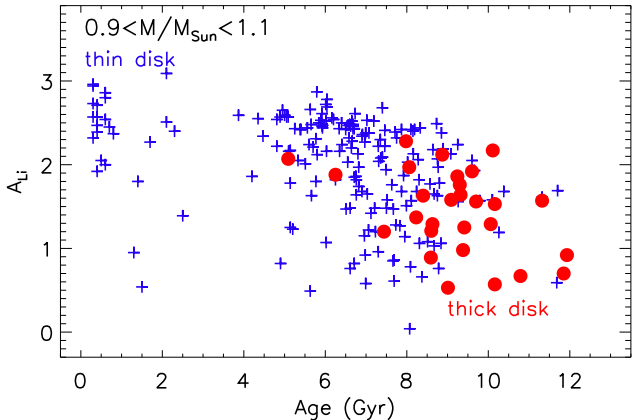}
\caption{Lithium abundance versus age for thin-disk (crosses) and thick-disk (circles) stars of mass $0.9<M/M_\odot<1.1$. 
Upper limits to lithium abundances based on the non detection of the lithium 6708\,\AA\ doublet are not shown.}
\label{f:gce3}
\end{figure}

Although the thick disk overlaps the thin disk in metallicity, it is systematically older than the thin disk and, therefore, the higher mass stars are found exclusively in the thin disk. For our first detailed comparison, we restrict the thin disk sample to stars with masses lower than $1.1\,M_\odot$ and then plot separately thin-disk stars older than 2, 5, and 8\,Gyr (Figure~\ref{f:gce2}). Previous studies have found that most thick-disk stars are older than 8\,Gyr \cite[][and references therein]{allende06}. The solid line in each of the four panels of Figure~\ref{f:gce2} denotes the upper envelope to the lithium abundance for thin-disk stars, i.e., stars younger than 2 Gyr fill in the gap between stars plotted in the $>$2\,Gyr panel and the solid line. The upper envelope sits progressively at lower $\ali$ for older stars, in fact matching the upper envelope for the thick disk for stars older than 8\,Gyr. Even though this is not entirely clear from an inspection of the two bottom panels of Figure~\ref{f:gce2}, we note that the few thin-disk stars with $\ali\simeq2.4$ are among the most massive stars shown in that panel, while the thick-disk stars with the highest $\ali$ have an average mass about 0.1\,$M_\odot$ lower compared to those thin-disk objects. When this subtle mass bias is removed, the transition of the $\ali$ upper envelope from old thin disk to thick disk is very smooth (see also below the discussion on Figure~\ref{f:gce3}).

For the three thin disk panels shown in Figure~\ref{f:gce2}, the $\ali$ versus [Fe/H] relations are similar. This similarity is consistent with the contention by \cite{sestito05} that lithium depletion is ineffective beyond an age of 1--2 Gyr, although this may apply only to the stars defining the upper envelope of the $\ali$ versus $\feh$ relation; Figure~\ref{f:li_age_dissect} suggests that lithium depletes throughout the main-sequence life of a $1\,M_\odot$, solar-metallicity star.

Figure~\ref{f:gce2} shows that over the [Fe/H] range for the thin disk, the upper envelope of the $\ali$ data is almost independent of [Fe/H], at a level that decreases only very slightly as the age limit is increased from 2 to 5 and then 8 Gyr. This maximum lithium abundance approaches the maximum $\ali$ shown by thick-disk stars over their full [Fe/H] range which extends to lower [Fe/H] than do the thin-disk stars. 

A second and related comparison of thin and thick disk is presented in Figure~\ref{f:gce3}, which shows lithium abundance versus age for thin- and thick-disk stars with $M=1.0\pm0.1\,M_\odot$. The mass interval chosen for this comparison ensures that relatively massive thin-disk stars, which have suffered little lithium depletion, and relatively low-mass thick-disk stars, which have endured severe lithium depletion, are not compared to each other. This mass range also contains an important number of both thin- and thick-disk stars, allowing a better comparative analysis. Figure~\ref{f:gce3} confirms the impression from Figure~\ref{f:gce2} that over the age range of thick-disk stars there is very little difference in the distribution of lithium abundances for the two disks. In fact, this thick-disk star $\ali$ versus age trend, particularly at ages older than about 8\,Gyr, appears to be a natural continuation of the thin disk sample. The fact that the few relatively young thick-disk stars (age$<8$\,Gyr) have somewhat low $\ali$ probably owes to the fact that they are among the most massive of this group of stars.

\subsubsection{A Primordial Lithium Abundance}

\begin{figure}
\includegraphics[bb=80 370 390 570,width=8.8cm]
{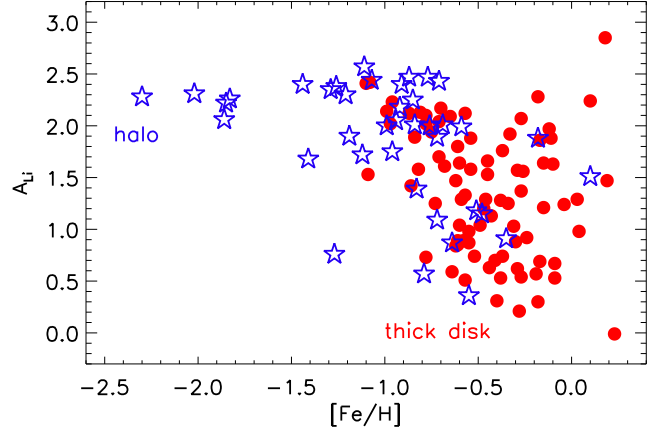}
\caption{Lithium abundance versus $\feh$ for our kinematically-selected thick-disk (filled circles) and halo stars (open symbols). Upper limits to lithium abundances based on the non detection of the lithium 6708\,\AA\ doublet are not shown.}
\label{f:gce4}
\end{figure}

Figure~\ref{f:gce4} shows the lithium abundances as a function of $\feh$ for our kinematically-selected thick-disk stars along with those derived for the halo stars in our sample. The maximum $\ali$ of the most metal-poor stars is comparable with that of the most metal-rich thick-disk stars. The star with very high $\ali\simeq2.8$ at $\feh\simeq0.2$ is too young to be a ``normal'' thick-disk member, as argued before (see footnote~\ref{tkta}). The observation that the maximum lithium abundance of thick-disk stars is equivalent to that in the more metal-poor stars on the Spite plateau recalls the question about whether the Spite plateau represents the primordial lithium abundance. If the answer is ``yes,'' the puzzle is then to show how the thick disk was able to increase its metallicity to [Fe/H] $\simeq -0.1$ without a significant increase in lithium abundance.\footnote{Note, however, that most of the iron in the thick disk was produced by Type~II supernovae whereas lithium originates from spallation in the interstellar medium, which could be a much slower process and have no significant production over the relatively short period when the thick disk was formed.} On the other hand, if the answer is ``no,'' the inference that the initial lithium abundance was that predicted for the {\it WMAP} baryon-to-photon ratio (i.e., $\ali\simeq2.7$) leads to the conclusion that important surface lithium depletions have occurred across the entire metallicity and mass range spanned by thick-disk and very metal-poor stars. 

\begin{figure}
\includegraphics[bb=85 375 510 650,width=9.0cm]
{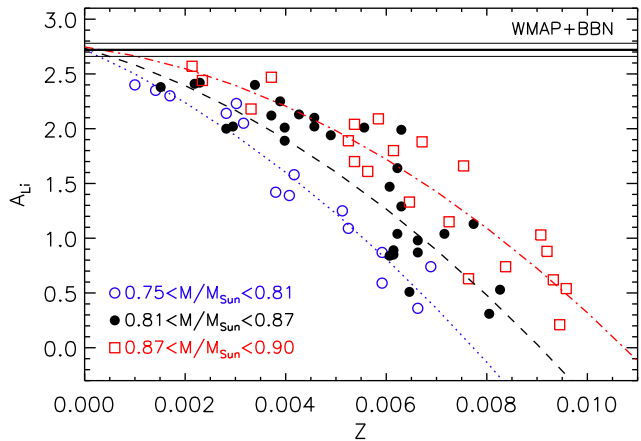}
\caption{Lithium abundance as a function of fractional metallicity for the kinematically-selected halo and thick-disk stars of our sample. Three mass groups are plotted, as described by the lower left legend. Quadratic fits to the lowest (dotted line) and highest (dash-dotted line) mass groups are overplotted, along with their average (dashed line). The thick and thin solid lines at $\ali=2.72\pm0.06$ correspond to the primordial lithium abundance inferred from \textit{WMAP} observations and standard Big Bang nucleosynthesis.}
\label{f:gce_z0}
\end{figure}

In Figure~\ref{f:gce_z0}, the lithium abundances of thick-disk and halo stars from our sample are plotted against their fractional metallicity $Z=Z_\odot\times10^\feh(0.638\times10^{[\alpha/\mathrm{Fe}]}+0.362)$, where $Z_\odot=0.014$ corresponds to the Sun, as given by \cite{asplund09:review}. Note that this formula for $Z$, which was derived by \cite{salaris93}, takes into account the impact of $\alpha$-element enhancement at low [Fe/H]. In our case, we assume $[\alpha/\mathrm{Fe}]=+0.4$ for $\feh<-0.7$ and linearly decreasing $[\alpha/\mathrm{Fe}]$ values from +0.4 at $\feh=-0.7$ to solar at $\feh=0$. The stars in Figure~\ref{f:gce_z0} have been sorted into three mass groups, as described by the legend on the lower left part of the Figure. Clearly, $\ali$ in these old stars depends on $Z$, but the relation is mass-dependent. At a given $Z$, lower mass stars have lower $\ali$.

The mass-dependent $\ali$ versus $Z$ relations for thick-disk and halo stars can be approximated with quadratic fits. Least-squares fits were applied to the lowest and highest mass groups, and are shown with the dotted and dot-dashed lines in Figure~\ref{f:gce_z0}. The dashed line corresponds to the average of these two quadratic fits, and it is a very good approximation to the $\ali$ versus $Z$ relation of the intermediate mass group. Interestingly, extrapolating these relations to $Z=0$ leads to a ``primordial'' lithium abundance $\ali\simeq2.73$, a value that is in excellent agreement with that derived from \textit{WMAP} observations and standard Big Bang nucleosynthesis $\ali=2.72\pm0.06$ \cite[e.g.,][]{coc12}.

Thus, the data for old thick-disk and halo stars reveals a lithium-mass-metallicity dependence that hints at a solution to the primordial lithium abundance problem. We note, however, that the physics behind this observation is likely related to the extent of the convective zone and its dependence on the stars' fundamental properties, whereas the problem with the Spite plateau requires additional depletion mechanisms, hence different physics. The fact that our extrapolation to $Z=0$ results in a high primordial lithium abundance consistent with the cosmological value could therefore be fortuitous.

Indeed, the Spite plateau is defined mostly by stars with $Z<0.002$, which are scarce in our work. Since nearly all of those objects have $\ali\simeq2.2$, the quadratic fits shown in Figure~\ref{f:gce_z0} may not apply to them. Perhaps the real $\ali$ versus $Z$ relation turns down towards lower $\ali$ for $Z<0.001$, leading to a low derived primordial lithium abundance.\footnote{In fact, in a very recent work, \cite{nissen12}, who perform a lithium-mass-metallicity fit and extrapolation to $Z=0$ nearly identical to ours, but with a linear relation, find that their fit using metal-rich halo stars and thick-disk stars does not reproduce well the very metal-poor star lithium abundance data by \cite{melendez10:lithium}.} However, if the masses of these very low metallicity stars (in our mass scale) turn out to be lower than those of the lowest mass group in our sample, they may exhibit a $Z$-dependence similar to that of the higher mass groups, implying a high primordial lithium abundance. We encourage future works on lithium abundances of large samples of very metal-poor stars to examine this relation carefully, including the more metal-rich halo and thick-disk stars.

\subsection{Lithium and Planets} 

The connection between stellar surface lithium abundance and presence of planets is the subject of intense debate. A number of authors claim to have found evidence for enhanced lithium depletion in planet-hosting stars \cite[e.g.,][]{chen06,israelian09,gonzalez10:lithium}, but others have concluded that the sensitivity of lithium abundance to other stellar parameters such as age and metallicity are responsible for the lithium abundance differences between stars with and without planets, which are therefore the product of sample biases and not related to the presence of planets \cite[e.g.,][]{ryan00,luck06,baumann10,ghezzi10}.

\begin{figure*}
\includegraphics[bb=70 360 910 1195,width=18.4cm]
{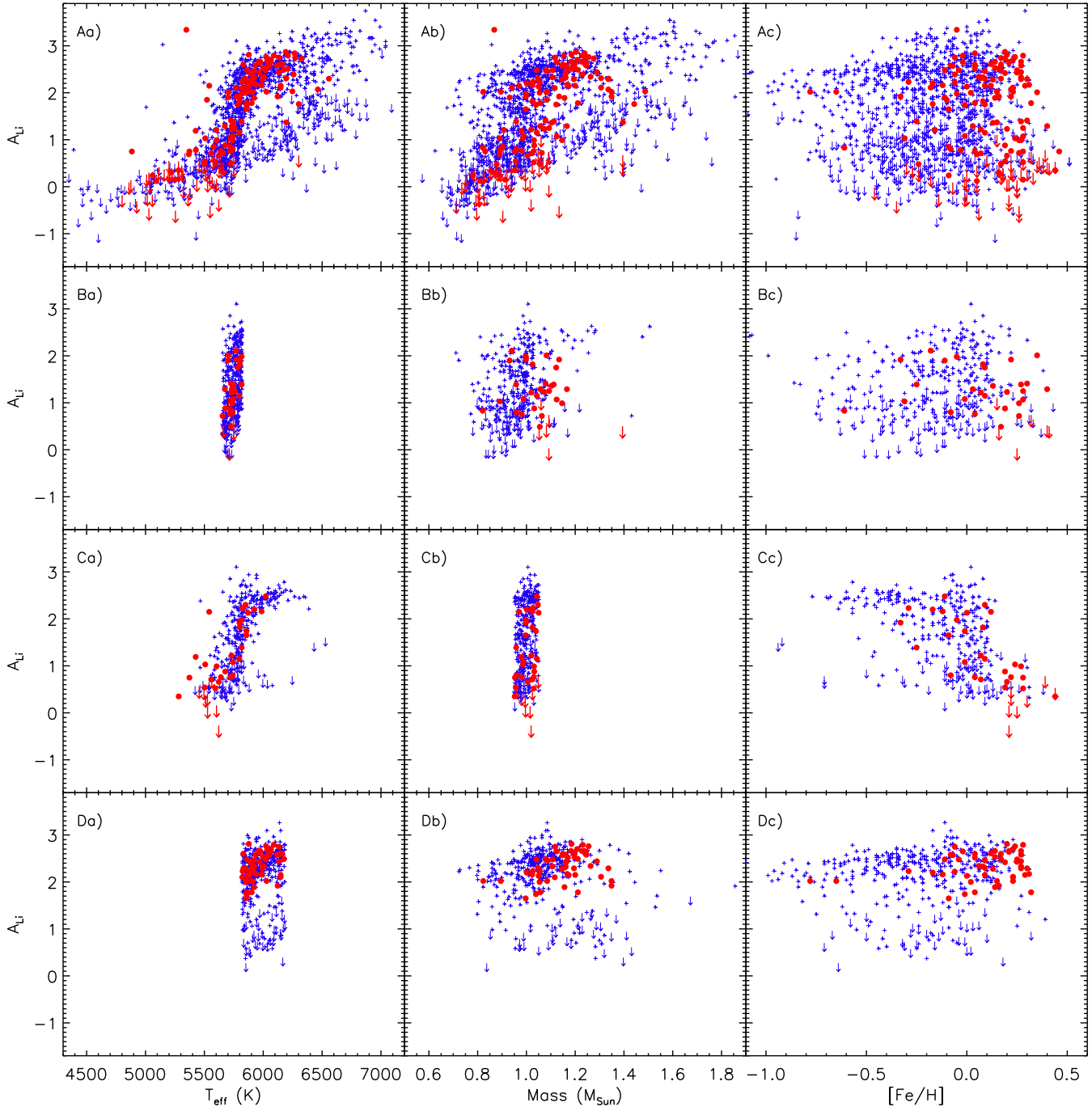}
\caption{Lithium abundance as a function of effective temperature (column a), mass (column b), and metallicity (column c) of the star. Known exoplanet hosts are shown with filled circles. Crosses correspond to stars for which no sub-stellar mass companions have been found yet, including stars which have not been searched for planets. Row A shows the entire sample. Row B shows only stars in the effective temperature range from 5650 to 5820\,K. In row C, only stars with a mass of $M=1.00\pm0.05\,M_\odot$ are shown. Row D shows only stars in the effective temperature range from 5820 to 6190\,K, i.e., in the $\teff$ range of the lithium desert. }
\label{f:li_planets}
\end{figure*}

There are 165 stars in our sample around which planets have been detected. Their lithium abundance patterns are shown in Figure~\ref{f:li_planets} along with those corresponding to the rest of our sample stars, i.e., to stars around which planets have not yet been detected. In this Figure, the comparison sample of non planet-hosts includes stars that have not been searched for sub-stellar mass companions. The heterogeneous nature of our sample regarding planet detection prevents us from making a more consistent planet/non-planet comparison. Lists of stars with and without planets can be found in the literature, but all one can say about the non planet-hosts is that they do not have planets below a certain mass and orbital period combination which depends on the instrumentation and observing strategy used by each group of investigators. Taking this into consideration, we prefer to compare our sample of known planet-hosts to the sample of otherwise ordinary field stars, some of which have been searched for planets.

Other than the lack of planet-hosts on the low lithium abundance side of the lithium desert (which is discussed below), a simple inspection of Figure~\ref{f:li_planets}Aa shows that the planet-host stars follow a very similar overall lithium abundance pattern compared to the sample of field stars. Planets are difficult to detect around warm, massive stars, which explains the lack planet-hosts on the hot $\teff$ and high mass ends of Figures~\ref{f:li_planets}Aa and \ref{f:li_planets}Ab. On the other hand, the planet-metallicity correlation, a very well defined trend at least for short-period massive planets around the dwarf stars which dominate our sample \cite[e.g.,][]{santos04,fischer05}, is behind the separation of samples in Figure~\ref{f:li_planets}Ac. It is interesting to notice that the planet-host sample defines a lithium dip similar to that observed in the Hyades (Figure~\ref{f:li_planets}Ab) better than the comparison sample. The fact that the planet-host sample has limited spread in [Fe/H] with a mean metallicity that is super-solar and the Hyades have $\feh\simeq+0.1$ while the comparison sample covers a very wide $\feh$ range could be the reason for this observation.

In order to better understand lithium depletion in planet hosts, it is customary to restrict the sample to a narrow range of effective temperature or mass. This practice, however, can be risky. To illustrate this, in Figure~\ref{f:li_planets}B we show the lithium abundance patterns of stars in the narrow $\teff$ range from 5650 to 5820\,K. A simple look at Figure~\ref{f:li_planets}Ba suggests that planet-hosts have systematically slightly lower lithium abundances. There are no planet-hosts with $\ali\gtrsim2$ in this effective temperature range, but a good number of comparison stars reside there. Closer inspection of this sample, however, allows us to better understand this trend and to realize that planets are not responsible for it. Figure~\ref{f:li_planets}Bb shows that the planet-host sample is on average more massive than the comparison sample. Although one could expect these stars to suffer less lithium depletion due to their thinner convective envelopes, we must notice that these objects, on average, are also more metal-rich than the comparison sample, as shown in Figure~\ref{f:li_planets}Bc. The majority of non planet-hosting stars with $\ali\gtrsim2$ in Figure~\ref{f:li_planets}B are either more massive than $M\simeq1.1\,M_\odot$ or more metal-poor than $\feh\simeq-0.1$. Both conditions imply higher lithium abundance, as shown by Figure~\ref{f:li_teff_dissect}.

Selecting stars within a narrow mass range does not help understanding the connection between lithium and planets either. For example, Figure~\ref{f:li_planets}Ca shows that when stars of nearly one-solar mass are selected $(M=1.00\pm0.05\,M_\odot)$, the range of $\teff$ covered by the sample is very large, owing to the different ages and metallicities of the otherwise ``Sun-like'' stars. As with the $\teff$ selection, this constraint in mass seems to also imply enhanced lithium depletion in planet-host stars. Figure~\ref{f:li_planets}Cb would suggest that the planet-host sample is shifted downwards in $\ali$ by about 0.5\,dex relative to the comparison stars. However, it is clear that this trend is due to the inclusion of a good number of relatively cool planet hosts with relatively massive convective envelopes, which could be slightly evolved $1\,M_\odot$ stars that have had plenty of time to deplete surface lithium. In addition, the planet-host star sample is again biased towards higher metallicities, which implies more lithium depletion at any given stellar age for stars of about one solar mass, as shown in Figure~\ref{f:li_age_dissect}. Figure~\ref{f:li_planets}Cc shows in fact a very clear correlation between lithium abundance and $\feh$ for stars of one solar mass, a correlation that is independent of whether the star has a planet or not.

Thus, the analysis of our large sample indicates that previously claimed patterns of enhanced lithium depletion in planet-host stars are a product of sample biases in stellar mass, age, and metallicity. \cite{baumann10} arrived at the exact same conclusion in their study of Sun-like stars. A system of particular interest in this context is the binary pair 16\,Cyg, which consists of two solar analog stars, one of which, the secondary, is known to host a gas giant planet and is also lithium poor relative to the primary, which appears to not have sub-stellar mass companions \cite[e.g.,][]{cochran97,king97}. Using our lithium abundance catalog, however, \cite{ramirez11} have shown that the lithium abundance difference seen in the 16\,Cyg pair can be fully explained by the slightly different masses and metallicities of the stars, leaving little or no room for a possible lithium--planet connection (see their Figure~11).

The only region of stellar parameter space that seems to separate planet hosts and comparison stars into high lithium and low lithium abundance samples is the region of the lithium desert (Figure~\ref{f:li_planets}Da). Interestingly, in this region ($5820<\teff<6190$\,K), the planet-hosts seem to have {\it higher} lithium abundance than ordinary field stars. More specifically, there are no planet-hosts on the low lithium abundance side of the lithium desert. Mass and metallicity effects cannot fully explain what is observed here. The mass distributions of the two samples are very similar (Figure~\ref{f:li_planets}Db) and, even though planet-hosts are, as usual, more metal-rich than the comparison stars, in the region of $\feh$ where there is significant overlap ($-0.1<\feh<+0.1$), one can still see that there are comparison stars with low lithium abundances (Figure~\ref{f:li_planets}Dc). Comparison stars with low lithium abundances in the desert region are still present if we restrict the sample to $\feh>-0.2$ and $M<1.2\,M_\odot$, i.e., when we make the two samples as similar as possible in their atmospheric parameters. A number of them (six) still remain there if we further restrict the comparison sample to stars that have been searched for planets but none have been detected around them yet. As explained in Sect.~\ref{s:desert}, stars on the low lithium abundance side of the desert are older than about 2\,Gyr. Only one of the planet-hosts in the $\teff$ range of the desert is younger than 2\,Gyr. Thus, the fact that the planet-host stars in Figure~\ref{f:li_planets}D have lithium abundances at least about one order of magnitude larger than those on the low lithium abundance side of the desert appears not to be due to an age bias either.

Our data suggest that some mechanism of planet-star interaction (not necessarily related to planet formation) prevents the star from undergoing the sudden surface lithium abundance drop responsible for the lithium desert. Nevertheless, we must practice extreme caution with this type of statement. Stars on the low side of the desert are relatively rare. Also, most of the planets discovered around our sample stars are gas giants. The existence of smaller planets around the stars on the low lithium abundance side of the lithium desert must not be ruled out. A systematic search for planets around the stars in the lithium desert region would be interesting to determine whether or not lithium abundances can really discriminate planet-hosts from non planet-host stars, at least within this narrow $\teff$ window. 

\section{CONCLUSIONS}

We have measured stellar parameters and lithium abundances of 671 stars, of which, to the best of our knowledge, 319 have their lithium abundances derived for the first time. Then, using data from the literature, we have constructed a catalog of stellar parameters, lithium abundances, masses, and ages for 1381 nearby FGK dwarf and subgiant stars. These data are used to investigate stellar lithium depletion inside stars, the Galactic chemical evolution of lithium, and the impact of planet formation on the internal evolution of cool stars.

A lithium desert is observed in the $\ali$ versus $\teff$ plane from about $\teff=5950$ to 6100\,K. Stars are found above and below $\ali\simeq1.8$, creating a gap of about 0.5\,dex in $\ali$. Many of the stars on the low lithium abundance side of the desert have only upper limits to $\ali$, implying that the gap could be deeper. Detailed inspection of $\ali$ versus mass, age, and metallicity plots suggests that a short-lived process depletes lithium on stellar surfaces during their main-sequence or subgiant phases, at least for stars with masses between 1.1 and $1.3\,M_\odot$. Stars from the lithium dip, a feature observed in some young open clusters, which have now evolved and separated from their parent clusters, may contribute to the low $\ali$ side of the lithium desert, but they are unlikely to be the main source of stars in that region of the $\ali$ versus $\teff$ plane.

We have found evidence of lithium depletion on the main sequence for stars other than the Sun, and have revealed a dependence of lithium depletion on the stellar metallicity, which is, however, not straightforward. Metal-rich stars deplete lithium quicker than metal-poor stars for $M<1.1\,M_\odot$, but the opposite is true for higher masses. Time-scales of main-sequence lithium depletion depend both on mass and metallicity, but not in a simple way.

We have attempted to detect, for the first time to the best of our knowledge, a difference in lithium abundances for thin- and thick-disk stars. However, the stars with maximum lithium abundance in each of these groups, i.e., those with the least amount of lithium depletion, have different fundamental properties. Thin-disk stars tend to be younger and more massive than thick-disk stars, leading to thin/thick disk lithium abundance differences that reflect different degrees of lithium depletion rather than differences in lithium enrichment of the interstellar medium. Nevertheless, we find that the maximum lithium abundance of thick-disk stars is very similar to the Spite plateau value defined by very old halo stars, suggesting that it may extend up to $\feh=-0.1$.

Combining thick-disk and halo stars, we find a well-defined relation between lithium abundance and fractional metallicity, which is mass-dependent. Extrapolated to zero metallicity, this relation implies a primordial lithium abundance of $\ali=2.73$, i.e., a value that is consistent with \textit{WMAP} observations and standard Big Bang nucleosynthesis. Although this hints at a solution to the primordial lithium abundance problem, we note that our work does not include many very metal-poor stars, which could affect the shape of the lithium-mass-metallicity relation, and therefore the primordial lithium abundance inferred from its extrapolation to zero metallicity.

Our investigation of lithium depletion in planet-host stars relative to ordinary field stars shows that a pattern of enhanced lithium depletion in planet-hosts can be created by restricting samples to narrow ranges of $\teff$ or mass. However, age and metallicity effects must be taken into account before drawing any conclusions from this biased pattern. When all relevant stellar parameters are properly taken into account, we find no evidence for enhanced lithium depletion in planet-hosts. In fact, we find instead that in the region of the lithium desert, no planet hosts are found with low lithium abundances. Only a dedicated search for planets around stars in the lower side of the lithium desert can tell us whether this is something related to a planet-star interaction or if it is due to sample biases, but with the data in hand the former scenario is favored.

\acknowledgments

I.R.'s work was performed under contract with the California Institute of Technology (Caltech) funded by NASA through the Sagan Fellowship Program. J.R.F. acknowledges support from the summer student program at Carnegie Observatories. D.L.L. wishes to thank the Robert\ A.\ Welch Foundation of Houston, Texas for support through grant F-634. The authors thank the referee, Dr.\ Luca Sbordone, for his suggestions to improve our manuscript.

\end{document}